\documentclass[preprint,11pt]{aastex}

\providecommand{\etal}{et~al.}
\providecommand{\ie}{i.e.}

\received{}
\revised{}
\accepted{}

\slugcomment{\aj\ in press}

\shorttitle{Optical Photometry of GRB~021004}
\shortauthors{S.~T. Holland, et~al.}

\begin{document}

%%%%%%%%%% Title & Author %%%%%%%%%%%%%%%%%%%%%%%%%%%%%%%%%%%%%%%%%%%%%%%%%%%%%%%

\title{Optical Photometry of GRB~021004: The First
            Month\protect\footnote{Based on observations taken with
            the Nordic Optical Telescope (NOT), operated on the island
            of Santa Miguel de la Palma jointly by Denmark, Finland,
            Iceland, Norway, and Sweden, in the Spanish Observatorio
            del Roque de los Muchachos of the Instituto de
            Astrof{\'\i}sica de Canarias, and on observations taken
            with the {\sl Chandra\/} $X$-Ray Observatory.}}

\author{Stephen~T.~Holland\altaffilmark{2},
        Michael~Weidinger\altaffilmark{3},
        Johan~P.~U.~Fynbo\altaffilmark{3},
        Javier~Gorosabel\altaffilmark{4,5},
        Jens~Hjorth\altaffilmark{6},
        Kristian~Pedersen\altaffilmark{6},
	Javier~M{\'e}ndez Alvarez\altaffilmark{11,12},
	Thomas~Augusteijn\altaffilmark{7},
	J.~M.~Castro~Cer{\'o}n\altaffilmark{8},
	Alberto~Castro-Tirado\altaffilmark{5},
        H{\aa}kon~Dahle\altaffilmark{9},
        M.~P.~Egholm\altaffilmark{3,7}, 
        P{\'a}ll~Jakobsson\altaffilmark{6},
        Brian~L.~Jensen\altaffilmark{6},
	Andrew~Levan\altaffilmark{10},
        Palle~M{\o}ller\altaffilmark{13},
        Holger~Pedersen\altaffilmark{6},
        Tapio~Pursimo\altaffilmark{7},
        Pilar~Ruiz-Lapuente\altaffilmark{12}, \&
        Bjarne~Thomsen\altaffilmark{3}}

\altaffiltext{2}{Department of Physics,
                 University of Notre Dame,
                 Notre Dame, IN 46556--5670,
                 U.S.A.
                 \email{sholland@nd.edu}}

\altaffiltext{3}{Department of Physics \& Astronomy,
                 University of Aarhus,
                 Ny Munkegade,
                 DK--8000 {\AA}rhus C,
                 Denmark
                 \email{michaelw@phys.au.dk,
                        jfynbo@phys.au.dk,
                        mpe@not.iac.es,
                        bt@phys.au.dk}}

\altaffiltext{4}{Laboratorio de Astrof{\'\i}sica Espacial y F{\'\i}sica Fundamental (INTA),
                 Apartado de Correos, 50.727,
                 E--28.080 Madrid,
                 Spain
                 \email{jgu@laeff.esa.es}}

\altaffiltext{5}{Instituto de Astrof{\'\i}sica de Andaluc{\'\i}a (IAA-CSIC),
                 Apartado de Correos, 3.004,
                 E--18.700 Granada,
                 Spain
                 \email{jgu@iaa.es,,
                        ajct@iaa.es}}

\altaffiltext{6}{Astronomical Observatory,
                 University of Copenhagen,
                 Juliane Maries Vej 30,
                 DK--2100 Copenhagen {\O},
                 Denmark
                 \email{jens@astro.ku.dk,
		        kp@astro.ku.dk,
		        pallja@astro.ku.dk,
                        brian\_j@astro.ku.dk,
                        holger@astro.ku.dk}}

\altaffiltext{7}{Nordic Optical Telescope,
                 Apartado de Correos, 474,
                 E--38.700 Santa Cruz de la Palma (Tenerife),
                 Spain
                 \email{tau@not.iac.es,
                        tpursimo@not.iac.es}}

\altaffiltext{8}{Real Instituto y Observatorio de la Armada,
                 Secci{\'o}n de Astronom{\'\i}a,
                 E--11.110 San Fernando--Naval  (C{\'a}diz),
                 Spain
                 \email{josemari@alumni.nd.ed}}

\altaffiltext{9}{NORDITA,
                 Blegdamsvej 17,
                 DK--2100 Copenhagen {\O},
                 Denmark
                 \email{dahle@nordita.dk}}

\altaffiltext{10}{Department of Physics and Astronomy,
                  University of Leicester,
                  University Road, Leicester, LE1 7RH,
                  UK
		  \email{anl@star.le.ac.uk}}   

\altaffiltext{11}{Isaac Newton Group of Telescopes,
                  Apartado de Correos 321,
                  E--38.700 Santa Cruz de la Palma,
                  Spain
                  \email{jma@ing.iac.es}}

\altaffiltext{12}{Departamento de Astronom{\'\i}a y Meteorolog{\'\i}a,
                  Universidad de Barcelona,
                  Mart{\'\i} i Franqu{\`e}s 1,
                  E--08.028 Barcelona,
                  Spain
                  \email{pilar@mizar.am.ub.es}}

\altaffiltext{13}{European Southern Observatory,
                  Karl--Schwarzschild--Stra{\ss}e 2,
                  D--85748 Garching bei M{\"u}nchen,
                  Germany
                  \email{pmoller@eso.org}}

%%%%%%%%%% Abstract %%%%%%%%%%%%%%%%%%%%%%%%%%%%%%%%%%%%%%%%%%%%%%%%%%%%%%%%%%%%

\begin{abstract}

     We present $U\!BV\!{R_C}{I_C}$ photometry of the optical
afterglow of the gamma-ray burst \objectname{GRB~021004} taken at the
Nordic Optical Telescope between approximately eight hours and 30 days
after the burst.  This data is combined with an analysis of the 87
ksec {\sl Chandra\/} $X$-ray observations of \objectname{GRB~021004}
taken at a mean epoch of 33~hours after the burst to investigate the
nature of this GRB\@.  We find an intrinsic spectral slope at optical
wavelengths of $\beta_{U\!H} = 0.39 \pm 0.12$ and an $X$-ray slope of
$\beta_X = 0.94 \pm 0.03$.  There is no evidence for color evolution
between 8.5 hours and 5.5 days after the burst.  The optical decay
becomes steeper approximately five days after the burst.  This appears
to be a gradual break due to the onset of sideways expansion in a
collimated outflow.  Our data suggest that the extra-galactic
extinction along the line of sight to the burst is between $A_V
\approx 0.3$ and $A_V \approx 0.5$ and has an extinction law similar
to that of the Small Magellanic Cloud.  The optical and $X$-ray data
are consistent with a relativistic fireball with the shocked electrons
being in the slow cooling regime and having an electron index of $p =
1.9 \pm 0.1$.  The burst occurred in an ambient medium that is
homogeneous on scales larger than approximately $10^{18}$~cm but
inhomogeneous on smaller scales.  The mean particle density is similar
to what is seen for other bursts ($0.1 \lesssim n \lesssim 100$
cm$^{-3}$).  Our results support the idea that the brightening seen
approximately 0.1 days was due to interaction with a clumpy ambient
medium within $10^{17}$--$10^{18}$~cm of the progenitor.  The
agreement between the predicted optical decay and that observed
approximately ten minutes after the burst suggests that the physical
mechanism controlling the observed flux at $t \approx 10$ minutes is
the same as the one operating at $t > 0.5$ days.

\end{abstract}

%%%%%%%%%% Keywords %%%%%%%%%%%%%%%%%%%%%%%%%%%%%%%%%%%%%%%%%%%%%%%%%%%%%%%%%%%%

\keywords{gamma rays: bursts}

%%%%%%%%%%%%%%%%%%%%%%%%%%%%%%%%%%%%%%%%%%%%%%%%%%%%%%%%%%%%%%%%%%%%%%%%%%%%%%%%

\section{Introduction\label{SECTION:intro}}

     The gamma-ray burst (GRB) \objectname{GRB~021004} was detected in
the constellation Pisces by the FREGATE, WXM, and SXC instruments on
board the {\sl High Energy Transient Explorer II\/} ({\sl HETE-II\/})
satellite at 12:06:13.57 UT on 2002~Oct.~4 \citep{SGM2002}.  The burst
had a duration of approximately 100 s and consisted of two peaks
separated by approximately 25 s.  Each peak had a fast rise and
exponential decay profile and a power-law spectrum with a slope of
1.64 \citep{LRA2002}.  The FREGATE instrument on {\sl HETE-II\/}
measured a fluence of $1.3 \times 10^{-6}$~erg~cm$^{-2}$ between 50
and 300~keV and $3.2 \times 10^{-6}$~erg~cm$^{-2}$ between 7 and
400~keV.  The WXM fluence (2--25~keV) is $7.5 \times
10^{-7}$~erg~cm$^{-2}$ \citep{LRA2002}.  Therefore,
\objectname{GRB~021004} was an $X$-ray rich burst and fits into the
long--soft class of bursts \citep{KMF1993}.

     The redshift of the burst was initially constrained to be $z \ge
1.60$ based on the detection of a \ion{Mg}{2} absorption system in the
spectrum of the afterglow \citep{FBS2002}. \citet{CF2002} found
absorption and emission due to Ly$\alpha$ at $z=2.33$. \citet{MFH2002}
report on five absorption systems and confirm the presence of the
Ly$\alpha$ emission line, from which they derive a host galaxy
redshift of $z = 2.3351$. The GRB absorption system shows strong
similarities with associated QSO absorbers, {\ie}, an outflow velocity
of several 1000~km~s$^{-1}$, high ionization, and line-locking
\citep{SRW2002,SFI2002,MFH2002}.

     \citet{TKY2002} observed the location of the GRB 3.5 minutes
after the burst with the RIKEN automated telescope and found an upper
limit for the unfiltered magnitude of 13.6.  The optical afterglow
(OA) was identified 9.45 minutes after the {\sl HETE-II\/} trigger by
the 48-inch Oschin/NEAT robotic telescope \citep{F2002}.  The rapid
identification allowed for near-continuous monitoring of this burst's
OA\@.  It quickly became apparent that the OA was not following a
simple power-law decay but had rebrightened approximately 0.1 days
after the burst.  Further observations suggested that there were rapid
variation in the optical decay \citep{WBS2002} similar to those seen
in \objectname{GRB~011211} \citep{HSG2002,JHF2002}, and larger
deviations from a power law decay on scales of hours to a day
\citep{WBS2002,SBA2002}.

     The large variations in \objectname{GRB~021004}'s optical decay
are unusual, but not unheard of in GRB afterglows.
\objectname{GRB~970508} was the second GRB for which an OA was
identified.  This OA had a constant luminosity for approximately one
day after the burst then brightened by a factor of approximately six
before taking on the familiar power-law decay \citep{CGB1998,PJG1998}.
\citet{PMR1998} showed that this behavior could be explained if there
is an additional injection of energy into the external shock at later
times.  This additional energy could come from a shell of ejecta
moving more slowly than the main body of ejecta impacting on the
external medium.  \objectname{GRB~000301C} also exhibited significant
deviations from a broken power-law decay for several days after the
burst.  \citet{GLS2000} found that these deviations were consistent
with the OA being microlensed.  \citet{HSG2002} and \citet{JHF2002}
observed rapid variations in the optical decay of
\objectname{GRB~011211} which \citet{HSG2002} interpret as being due
to small-scale density fluctuations in the ambient medium within
0.1~pc of the progenitor.  As rapid responses to GRBs become more
common optical observations will be able to probe the first few hours
of more GRBs and allow observations of the prompt emission.  This will
provide a window into the physics of the early OA\@.  In this paper we
present a self-consistent set of optical observations of
\objectname{GRB~021004} starting approximately 8.5~hours after the
burst and covering the first month.  We use this data to constrain the
physics of the relativistic fireball and probe the ambient medium
around the progenitor.
     
     In this paper we adopt a cosmology with a Hubble parameter of
$H_0 = 65$~km~s$^{-1}$~Mpc$^{-1}$, a matter density of $\Omega_m =
0.3$, and a cosmological constant of $\Omega_\Lambda = 0.7$.  For this
cosmology a redshift of $z = 2.3351$ corresponds to a luminosity
distance of 20.22~Gpc and a distance modulus of 46.53.  One arcsecond
corresponds to 29.39 comoving kpc, or 8.81 proper kpc.  The lookback
time is 11.56~Gyr.

%%%%%%%%%%%%%%%%%%%%%%%%%%%%%%%%%%%%%%%%%%%%%%%%%%%%%%%%%%%%%%%%%%%%%%%%%%%%%%%%%

\section{The Data\label{SECTION:data}}

\subsection{Optical Photometry\label{SECTION:phot}}

     The OA for \objectname{GRB~021004} is located at R.A. =
00:26:54.69, Dec.\ = +18:55:41.3 (J2000) \citep{F2002}, which
corresponds to Galactic coordinates of $(b^{\mathrm{II}},
l^{\mathrm{II}}) = (-43\fdg5616, 114\fdg9172)$. The reddening maps of
\citet{SFD1998} give a Galactic reddening of $E_{B-V} = 0.060 \pm
0.020$~mag in this direction.  The corresponding Galactic extinctions
are $A_U = 0.325$, $A_B = 0.258$, $A_V = 0.195$, $A_{R_C} = 0.160$,
$A_{I_C} = 0.116$, and $A_H = 0.034$.
                                            
\begin{figure}[ht]
%\figurenum{}
%\epsscape{}
\plotone{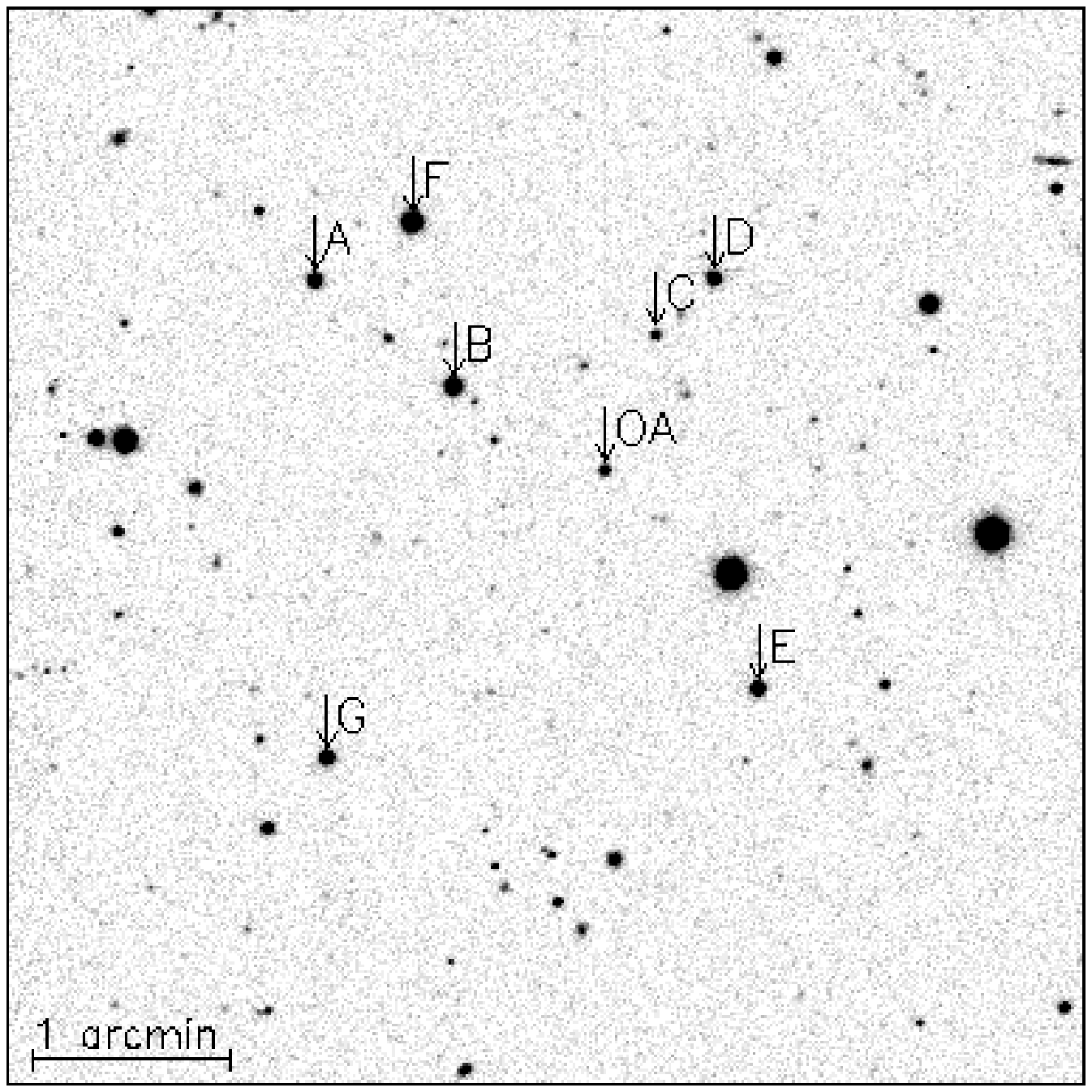}
\figcaption[./Holland.fig1.ps]{The field of \protect\objectname{GRB~021004}
in the $R_C$ band.  The stars marked A--G are used as secondary
standards for the relative photometry.  The positions and photometry
of these secondary standard stars are given in
Table~\ref{TABLE:standards}.  North is up and east is to the left.
The OA is isolated so there is no confusion with neighbouring
sources.\label{FIGURE:finder}}
\end{figure}

     We obtained $U\!BV\!{R_C}{I_C}$ images of the field containing
\objectname{GRB~021004} using the Andaluc{\'\i}a Faint Object
Spectrograph and Camera (ALFOSC) and the Mosaic Camera (MOSCA) on the
2.56-m Nordic Optical Telescope (NOT) at La Palma between 2002~Oct.~4
and 2002~Nov.~4.  The ALFOSC detector is a $2048 \times 2048$ pixel
thinned Loral CCD with a pixel scale of $0\farcs189$. The instrumental
gain was 1.0~e$^-$/ADU and the read-out noise was 6~e$^-$/pixel.  The
MOSCA detector is a $2 \times 2$ mosaic CCD camera containing four
flash-gated Loral-Lesser thinned $2048 \times 2048$ CCDs.  The pixel
size is 15~$\mu$m and the pixel scale is $0\farcs11$.  The
instrumental gain is 1.24~e$^-$/ADU and the read-out noise is
8.5~e$^-$/pixel.  The data from Oct.~19 and Nov.~4 were obtained using
MOSCA\@.  All other data were obtained using ALFOSC\@.
Fig.~\ref{FIGURE:finder} shows the field of \objectname{GRB~021004}.

%===== Begin Finding Chart Figure =====%
%===== End Finding Chart Figure =====%

     The data were preprocessed using standard techniques for bias and
flat-field corrections.  Photometry was performed using {\sc
DAOPhot~II} \citep{S1987,SH1988} and calibrated using seven secondary
standard stars from the catalogue of \citet{H2002} (see
Table~\ref{TABLE:standards}).  The magnitude of the OA at each epoch
was determined as the weighted average over the OA magnitudes computed
relative to each of the secondary standard stars.  The weights for
each point were the quadratic sum of the {\sc DAOPhot II} error and
the photometric error for each standard star.  The corresponding error
in the OA at each epoch was computed as the dispersion of the OA
magnitude calculated relative to each of the secondary standards.  The
log of the observations and the photometry of the OA is presented in
Table~\ref{TABLE:phot}.  In some images stars B and F were saturated
and were not used for calibration.  \citet{H2002} did not detect stars
C and D in the $U$-band, so we could not use them to calibrate our
$U$-band photometry.  The color terms in the calibrations were smaller
than 5\% and did not improve the quality of the calibrations.
Therefore no color corrections were applied to the photometry.

%===== Begin Calibration Star Table =====%
\begin{deluxetable}{cccccccc}
\tabletypesize{\scriptsize}
%\rotate
\tablewidth{0pt}
%\tablenum{}
%\tablecomumns{}
%\tableheadfrac{}
\tablecaption{Positions and photometry of the secondary standard stars
A--G taken from \protect\citet{H2002}.\label{TABLE:standards}}
\tablehead{%
     \colhead{Star} &
     \colhead{R.A.} &
     \colhead{Dec.} &
     \colhead{$U$} &
     \colhead{$B$} &
     \colhead{$V$} &
     \colhead{$R_C$} &
     \colhead{$I_C$}}
\startdata
A & 00:27:00.84 & +18:56:38.4 & $20.699 \pm 0.038$ & $19.480 \pm 0.018$ & $17.989 \pm 0.016$ & $17.058 \pm 0.023$ & $16.084 \pm 0.051$ \\
B & 00:26:57.89 & +18:56:06.7 & $17.372 \pm 0.027$ & $17.341 \pm 0.026$ & $16.699 \pm 0.006$ & $16.333 \pm 0.013$ & $15.948 \pm 0.025$ \\
C & 00:26:53.61 & +18:56:22.6 &       \nodata      & $21.116 \pm 0.11$  & $19.741 \pm 0.037$ & $18.785 \pm 0.057$ & $17.828 \pm 0.065$ \\
D & 00:26:52.36 & +18:56:39.5 &       \nodata      & $19.765 \pm 0.032$ & $18.287 \pm 0.018$ & $17.364 \pm 0.028$ & $16.455 \pm 0.033$ \\
E & 00:26:51.42 & +18:54:36.0 & $18.099\pm 0.021$  & $18.091 \pm 0.013$ & $17.493 \pm 0.009$ & $17.142 \pm 0.013$ & $16.787 \pm 0.027$ \\
F & 00:26:58.77 & +18:56:56.0 & $18.545 \pm 0.044$ & $17.430 \pm 0.029$ & $16.258 \pm 0.006$ & $15.538 \pm 0.016$ & $14.896 \pm 0.028$ \\
G & 00:27:00.56 & +18:54:14.8 & $17.616 \pm 0.060$ & $17.836 \pm 0.033$ & $17.323 \pm 0.006$ & $17.002 \pm 0.008$ & $16.643 \pm 0.015$ \\
\enddata
\end{deluxetable}
%===== End Calibration Star Table =====%

%===== Begin Table of Photometry =====%
\begin{deluxetable}{rccr|rccr}
\tabletypesize{\scriptsize}
%\rotate
\tablewidth{0pt}
%\tablenum{}
%\tablecomumns{}
%\tableheadfrac{}
\tablecaption{Log of the \protect\objectname{GRB~021004} observations
and the results of the photometry.  The UT date is the middle of each
exposure.\label{TABLE:phot}}
\tablehead{%
     \colhead{UT Date} &
     \colhead{Magnitude} &
     \colhead{Seeing ($\arcsec$)} &
     \colhead{$t$ (s)} &
     \colhead{UT Date} &
     \colhead{Magnitude} &
     \colhead{Seeing ($\arcsec$)} &
     \colhead{$t$ (s)}}
\startdata   
{\bf $U$-band:}&                 &     &      & {\bf $R_C$-band cont.:}&        &     & \\     
Oct.\  6.8880 & $20.88 \pm 0.02$ & 1.3 & 1000 &       6.0092 & $19.76 \pm 0.03$ & 1.2 &  300 \\
       7.8880 & $21.18 \pm 0.02$ & 1.9 & 1000 &       6.1221 & $19.91 \pm 0.03$ & 1.6 &  600 \\
       8.8880 & $21.52 \pm 0.03$ & 1.7 & 1000 &       6.1733 & $19.94 \pm 0.03$ & 1.4 &  600 \\
      10.0925 & $21.92 \pm 0.12$ & 1.8 &  900 &       6.2085 & $19.92 \pm 0.03$ & 1.6 &  600 \\
              &                  &     &      &       6.8348 & $20.03 \pm 0.03$ & 1.4 &  600 \\
{\bf $B$-band:}&                 &     &      &       6.8633 & $20.10 \pm 0.03$ & 1.3 &  600 \\
Oct.\  5.9441 & $20.71 \pm 0.02$ & 1.0 &  600 &       6.9175 & $20.13 \pm 0.03$ & 1.2 &  600 \\
      10.0639 & $22.15 \pm 0.05$ & 1.8 &  600 &       6.9255 & $20.12 \pm 0.03$ & 1.2 &  600 \\
              &                  &     &      &       7.0391 & $20.16 \pm 0.03$ & 1.0 &  600 \\
{\bf $V$-band:}&                 &     &      &       7.8451 & $20.40 \pm 0.03$ & 1.5 &  600 \\
Oct.\ 10.0710 & $21.49 \pm 0.09$ & 1.8 &  300 &       7.8531 & $20.40 \pm 0.03$ & 1.5 &  600 \\
              &                  &     &      &       7.8925 & $20.43 \pm 0.03$ & 1.5 &  600 \\
{\bf $R_C$-band:}&               &     &      &       7.9479 & $20.42 \pm 0.03$ & 1.5 &  600 \\
Oct.\  4.8472 & $18.02 \pm 0.03$ & 1.5 &  300 &       8.0151 & $20.43 \pm 0.03$ & 1.2 &  600 \\
       4.8517 & $18.04 \pm 0.03$ & 1.5 &  300 &       8.1412 & $20.49 \pm 0.03$ & 1.2 &  600 \\
       4.9103 & $18.23 \pm 0.03$ & 1.6 &  300 &       8.8514 & $20.70 \pm 0.03$ & 1.5 &  600 \\
       4.9445 & $18.32 \pm 0.03$ & 1.8 &  300 &       8.9249 & $20.77 \pm 0.03$ & 1.4 &  600 \\
       4.9987 & $18.60 \pm 0.03$ & 1.8 &  300 &       9.0047 & $20.81 \pm 0.03$ & 1.1 &  600 \\
       5.0638\tablenotemark{a} &
                $18.84 \pm 0.02$ & 3.6 &  300 &       9.0726 & $20.83 \pm 0.02$ & 1.5 &  600 \\
       5.0687\tablenotemark{a} &
                $18.86 \pm 0.03$ & 3.3 &  300 &       9.1434 & $20.85 \pm 0.03$ & 1.3 &  600 \\
       5.1193\tablenotemark{a} & 
                $18.98 \pm 0.02$ & 3.7 &  500 &      10.0759 & $21.05 \pm 0.02$ & 1.5  & 300 \\
       5.1264\tablenotemark{a} &
                $18.98 \pm 0.02$ & 3.3 &  500 &      10.1022 & $21.08 \pm 0.02$ & 1.5 & 6480 \\
       5.1332\tablenotemark{a} &
                $19.01 \pm 0.02$ & 4.1 &  500 &      14.0431 & $21.90 \pm 0.02$ & 0.8 &  600 \\
       5.1402\tablenotemark{a} &
                $19.02 \pm 0.03$ & 3.7 &  500 &      14.0518 & $21.95 \pm 0.03$ & 0.9 &  600 \\
       5.1470\tablenotemark{a} &
                $19.06 \pm 0.02$ & 3.6 &  500 &      14.0600 & $21.93 \pm 0.03$ & 0.9 &  600 \\
       5.1541\tablenotemark{a} &
                $19.08 \pm 0.03$ & 4.0 &  500 &      19.0622\tablenotemark{b}
                                                             & $22.70 \pm 0.16$ & 1.1 & 1800 \\
       5.1609\tablenotemark{a} &
                $19.09 \pm 0.02$ & 4.0 &  500 &      27.9970 & $23.22 \pm 0.08$ & 1.0 & 3600 \\
       5.1678\tablenotemark{a} &
                $19.13 \pm 0.02$ & 4.7 &  500 & Nov.\ 4.0158\tablenotemark{b}
                                                             & $23.54 \pm 0.11$ & 1.0 & 7200 \\
       5.1746\tablenotemark{a} &              
                $19.11 \pm 0.02$ & 5.1 &  500 &              &                  &     & \\
       5.1802\tablenotemark{a} &
                $19.11 \pm 0.02$ & 4.7 &  300 & {\bf $I_C$-band:}&              &     & \\
       5.1860\tablenotemark{a} &
                $19.13 \pm 0.02$ & 3.7 &  500 & Oct.\ 4.8563 & $17.55 \pm 0.02$ & 1.3 &  300 \\
       5.1929\tablenotemark{a} &
                $19.14 \pm 0.03$ & 3.8 &  500 &       7.8851 & $19.92 \pm 0.02$ & 1.2 &  500 \\
       5.8541 & $19.61 \pm 0.03$ & 2.0 &  600 &       8.8851 & $20.27 \pm 0.02$ & 1.5 &  500 \\
       5.8621 & $19.62 \pm 0.03$ & 1.7 &  600 &      10.0828 & $20.49 \pm 0.01$ & 1.3 &  600 \\
       5.9358 & $19.69 \pm 0.03$ & 1.8 &  600 \\
\enddata
\tablenotetext{a}{The image was rebinned $2\times 2$ to improve the
signal-to-noise ratio.}
\tablenotetext{b}{The image was obtained with MOSCA.}
\end{deluxetable}
%===== End Photometry Table =====%

%%%%%%%%%%%%%%%%%%%%%%%%%%%%%%%%%%%%%%%%%%%%%%%%%%%%%%%%%%%%%%%%%%%%%%%%%%%%%%%%%

\subsection{$X$-Ray Data\label{SECTION:xray}}

\begin{figure}
%\figurenum{}
%\epsscale{}
\plotone{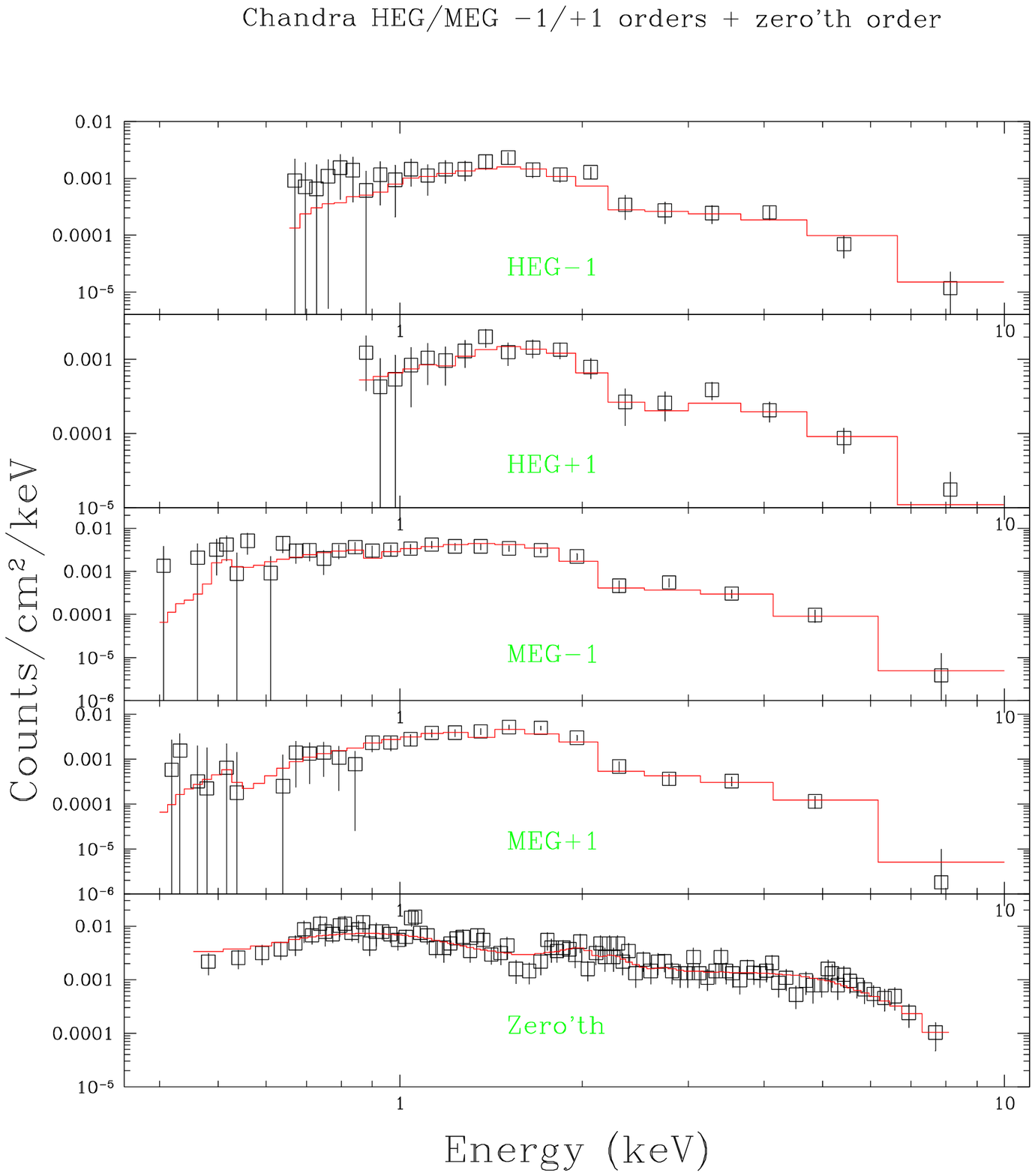}
\figcaption[./Holland.fig2.ps]{First order {\sl Chandra\/} High
Energy Grating/Medium Energy Grating spectra and zero'th order
spectrum of \protect\objectname{GRB~021004} with the best fit model
jointly fit to all spectra.  The lines show an intrinsic power law
with no absorption in the vicinity of \protect\objectname{GRB~021004}
(at $z=2.3351$) and a fixed Galactic absorption of $N(\mathrm{H}) =
1.23 \times 10^{20}$~cm$^{-2}$.\label{FIGURE:xray}}
\end{figure}

     {\sl Chandra\/} High-Energy Transmission Grating data were
obtained between 2002~Oct.~5.37 and 2002~Oct.~6.38 \citep{SH2002}.  We
analyzed the zero'th and first order spectra adopting standard
screening criteria for {\sl Chandra\/} data.  In order to constrain
the spectral index in the $X$-ray band, and to look into any
dependence on absorption, we made two joint zero'th and first order
fits to the data.  First, a power-law model was fit to the data in the
2--10~keV band.  Second, the data in the full-well calibrated spectral
range 0.4--10~keV was fit with an intrinsic power-law model with
absorption by Galactic hydrogen.  The column density was fixed at
$1.23 \times 10^{20}$~cm$^{-2}$ \citep{DL1990} while the absorbing
column at the redshift of \objectname{GRB~021004} was left as a free
parameter.  The {\sl Chandra\/} data and our fits are shown in
Fig.~\ref{FIGURE:xray}.

%===== Begin Chandra Figure =====%
%===== End Chandra Figure =====%

     A pure power law is an excellent fit ($\chi^2=47.8$ for 74
degrees of freedom (DOF)) to the data in the 2--10~keV band with a
best fit spectral index of $\beta_X = 1.03 \pm 0.06$.  A lower, but
consistent, spectral index ($\beta_X = 0.94 \pm 0.03$ with $\chi^2 =
127$ for 198 DOF) is obtained when including the absorption and
fitting data in the full 0.4--10~keV band (this data has superior
photon statistics compared to the 2--10~keV band).  We adopt $\beta_X
= 0.94 \pm 0.03$ for the $X$-ray spectral slope 1.4 days after the
burst since this value was derived from the spectrum with the better
noise properties.

     No extragalactic absorption is required along the line of sight
to \objectname{GRB~021004} and the upper limit on the absorbing column
density is $3.4 \times 10^{21}$~cm$^{-2}$.  This result is consistent
with the upper limit of $1.1 \times 10^{20}$~cm$^{-2}$ on the
\ion{H}{1} column density found by \citet{MFH2002}.  Scaling the value
of the fixed Galactic absorption up (down) by a factor of two results
in only a 0.03 decrease (increase) in the value of the best-fitting
spectral index.  We find no evidence for additional spectral features
such as lines or absorption edges.

%%%%%%%%%%%%%%%%%%%%%%%%%%%%%%%%%%%%%%%%%%%%%%%%%%%%%%%%%%%%%%%%%%%%%%%%%%%%%%%%%

\section{Results\label{SECTION:results}}

\subsection{The Slope of the Optical Spectrum\label{SECTION:extinction}}

     We used our $U\!BV\!{R_C}{I_C}$ photometry from Oct.~10 and the
near-simultaneous $H$-band photometry of \citet{SCM2002} to construct
the spectral energy distribution (SED) of the OA 5.568 days after the
burst.  The optical magnitudes were rescaled to the epoch of the
$H$-band measurement (Oct.\ 10.072 UT) and converted to flux densities
based on \citet{FSI1995}.  $H$-band magnitude--flux density conversion
factor was taken from \citet{A2000}.  Finally, the photometric points
were corrected for Galactic reddening.

     The SED was fit by $f_{\nu}(\nu) \propto \nu^{-\beta_{U\!H}}
\times 10^{-0.4 A(\nu)}$, where $f_\nu(\nu)$ is the flux density at
frequency $\nu$, $\beta_{U\!H}$ is the intrinsic optical spectral
index between the $U$ and $H$ photometric bands (approximately
3600--16\,000 {\AA}), and $A(\nu)$ is the extra-galactic extinction
along the line of sight to the burst.  The dependence of $A(\nu)$ on
$\nu$ has been parameterized in terms of the rest frame $A_V$
following the three extinction laws given by \citet{P1992} for the
Milky Way (MW), the Large Magellanic Cloud (LMC), and the Small
Magellanic Cloud (SMC).  The fit provides $\beta_{U\!H}$ and $A_V$
simultaneously for the assumed extinction laws.  For comparison
purposes the unextincted case ($A_V = 0$) was also considered.

\begin{figure}
%\figurenum{}
%\epsscale{}
\plotone{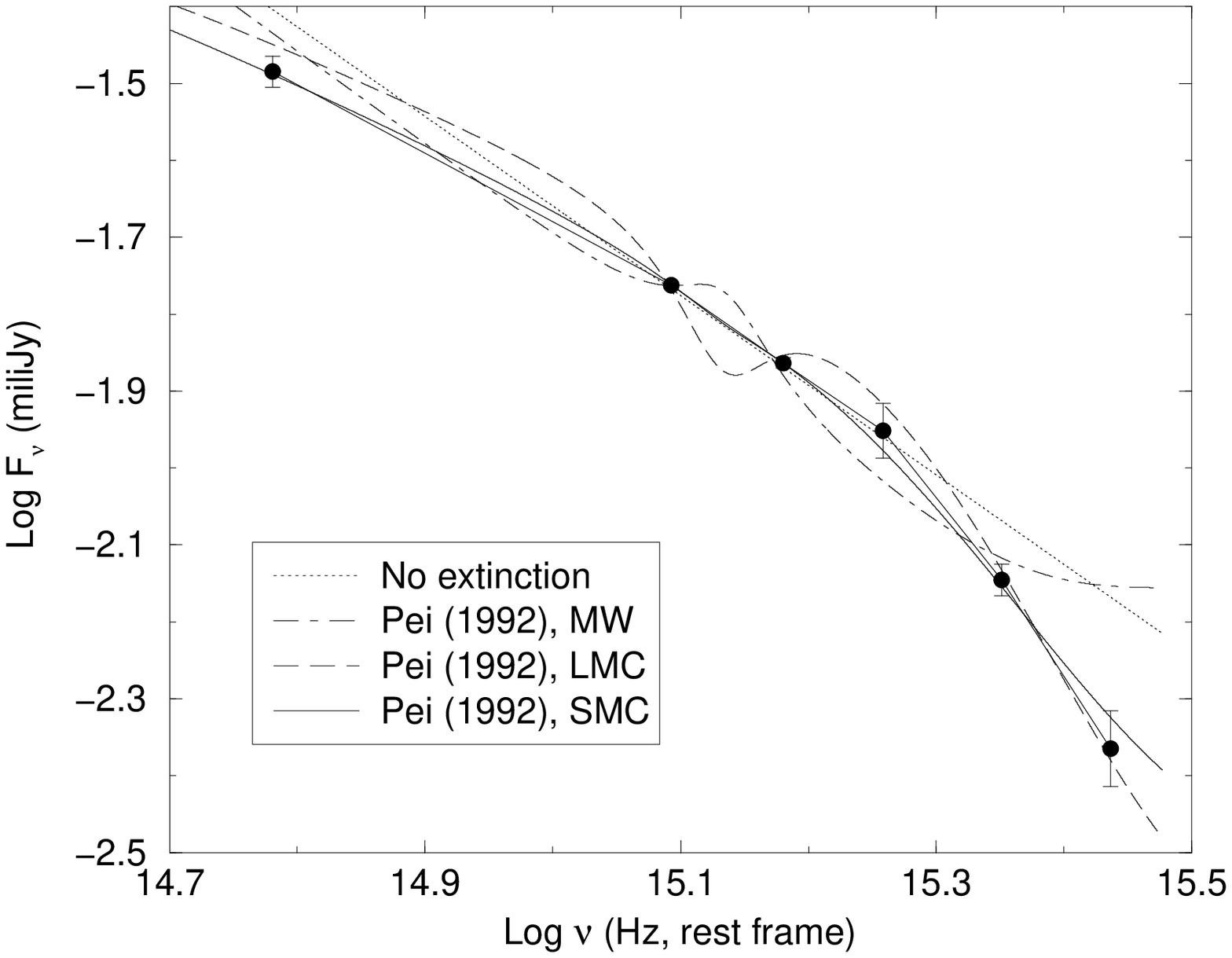}
\figcaption[./Holland.fig3.ps]{The SED of the OA of
\protect\objectname{GRB~021004} on 2002 Oct.\ 10.072 UT\@.  The filled
circles represent our NOT $U\!BV\!{R_C}{I_C}$ data and the $H$-band
data point from \citet{SCM2002}.  In the $R_C$ and $I_C$ bands the
error bars are smaller than the circles.  The lines represent the SED
fits when the SMC (solid), LMC (dashed) and MW (dot-dashed) extinction
laws given by \citet{P1992} are applied.  A fit (dotted line) assuming
no extinction of the host is shown for comparison.  If we assume that
the unextincted spectrum follows $f_\nu(\nu) \propto \nu^{-\beta}$
then only the SMC extinction provides an acceptable fit to the data
points.\label{FIGURE:sed}}
\end{figure}

     Fig.~\ref{FIGURE:sed} shows that $A_V = 0$ is not consistent with
our data ($\chi^2/\mathrm{DOF} = 12.3$).  We find the best fit
($\chi^2/\mathrm{DOF}=0.49$) with an SMC extinction law having $A_V =
0.26 \pm 0.04$ and $\beta_{U\!H} = 0.39 \pm 0.12$.  The MW and LMC
extinction laws give unacceptable fits ($\chi^2/\mathrm{DOF}=16.4$ and
$\chi^2/\mathrm{DOF}=3.8$ respectively).  This is the shallowest
spectral slope seen for any GRB OA\@.  We find no evidence that the
slope of the optical SED changed between approximately 0.35 and 5.5
days after the burst.  If we fix $\beta_{U\!H} = 1$, so that it is the
same as the spectral slope at $X$-ray frequencies, then none of the
extinction laws give acceptable fits ($\chi^2/\mathrm{DOF} > 8$).
This strongly suggests that there is a spectral break between the
optical and $X$-ray bands at $t = 1.4$ days.

%===== Begin SED Figure =====%
%===== End SED Figure =====%

     The red edge of the Ly$\alpha$ forest for a source at $z =
2.3351$ lies between the $U$ and $B$ bands, so our $U$-band data could
be significantly affected by intergalactic absorption.  In order to
test this we repeated our fits using only the $BV\!{R_C}{I_C}H$
photometry.  The results are consistent to within $1\sigma$ of those
obtained if the $U$-band data is included.  Therefore be believe that
intergalactic Ly$\alpha$ absorption is not significantly affecting our
estimate of the intrinsic spectral slope at optical wavelengths.

     \citet{MFH2002} find absorption systems at $z = 1.38$ and 1.60 in
addition to those at $z \approx 2.3$.  We repeated our fits placing
the absorbing dust at these lower redshifts and found that the fits
were comparable to those for absorption at $z = 2.3351$.  Therefore,
we are not able to constrain the redshift of the dust.  The total
absorption is $0.3 \lesssim A_V \lesssim 0.5$ in the rest frame of the
dust regardless of redshift of the dust.

%%%%%%%%%%%%%%%%%%%%%%%%%%%%%%%%%%%%%%%%%%%%%%%%%%%%%%%%%%%%%%%%%%%%%%%%%%%%%%%%%

\subsection{The Location of the Cooling Break\label{SECTION:break}}

     We find $U\!-\!R_C = 0.76 \pm 0.02$ between Oct.\ 6.9 and Oct.\
10.1.  When Galactic reddening is taken into consideration this color
is consistent with the Oct.~7.29 spectrum of \citet{MGF2002}.  This
suggests that there was no significant spectral evolution in the
optical between 2.4 and 5.5 days after the burst.  Our photometry also
shows no evidence for color evolution redward of approximately 6588
{\AA} between 0.35 and 5.5 days, therefore we believe that
$\beta_{U\!H} = 0.39 \pm 0.12$ at optical wavelengths between 0.35 and
5.5 days.  The lack of color evolution redward of approximately 6588
{\AA}, and the change in the intrinsic spectral slope between the
optical and $X$-ray bands, suggest that the increase in $B\!-\!V$ seen
by \citet{MGF2002} between 0.76 and 2.75 days after the burst was not
due to a spectral break passing through the optical frequencies.

     The differnt values for the optical and $X$-ray spectral slopes
1.4 days after the suggests that there is a spectral break between
them at this time.  We find $R_C\!-\!I_C = 0.53 \pm 0.02$ for $0.35 <
t < 5.5$ days, which suggests that this break did {\sl not\/} pass
through the optical during our observations.

     The relationships between $\beta_{U\!H}$, the slope of the
optical decay, $\alpha$, and the electron index, $p$, given by
\citet{SPH1999} (for a homogeneous interstellar medium (ISM)) and
\citet{CL1999} (for a pre-existing stellar wind) allow us to use
$\beta_{U\!H}$ and $\beta_X$ to predict $p$ and $\alpha$ during this
period.  These predictions are listed in Table~\ref{TABLE:predict}
where $p_{U\!H}$ and $p_X$ are the electron indices predicted from the
optical and $X$-ray spectral slopes respectively.  In cases where $1 <
p < 2$ the relationships of \citet{DC2001} were used.  Situations with
$p < 1$ are unphysical and can be ruled out.  We can also rule out
cases where both the cooling frequency, $\nu_c$, and the synchrotron
frequency, $\nu_m$, are above the optical since they predict a rising
spectrum ($\beta_{U\!H} < 0$) in the optical.  Therefore, our results
that $\beta_{U\!H} = 0.39 \pm 0.12$ and $\beta_X = 0.94 \pm 0.03$
require that either $\nu_m < \nu <\nu_c$ or $\nu_c < \nu < \nu_m$.

%===== Begin Table of Predictions =====%
\begin{deluxetable}{clccll}
\tabletypesize{\scriptsize}
%\rotate
\tablewidth{0pt}
%\tablenum{}
%\tablecomumns{}
%\tableheadfrac{}
\tablecaption{Predicted electron indices and decay slopes assuming
$\beta_{U\!H} = 0.39 \pm 0.12$ and $\beta_X = 0.94 \pm 0.03$.  The
predicted early-time slope of the optical decay is denoted
$\alpha$.  \label{TABLE:predict}}
\tablehead{%
        \colhead{Model} &
        \colhead{Environment} &
        \colhead{$p_{U\!H}$} &
        \colhead{$p_X$} &
        \colhead{$\alpha$} &
        \colhead{Comments}}
\startdata
 $\nu_m < \nu_c < \nu$ & ISM  & $0.8 \pm 0.2$ & $1.9 \pm 0.1$ & \nodata         & $p_{U\!H}$ and $p_X$ are inconsistent \\
                       & Wind &               &               & \nodata         & $p_{U\!H}$ and $p_X$ are inconsistent \\
 $\nu_m < \nu < \nu_c$ & ISM  & $1.8 \pm 0.2$ & $1.9 \pm 0.1$ & $0.73 \pm 0.02$ & \\
                       & Wind &               &               & $1.24 \pm 0.01$ & $\alpha$ does not fit data \\
 $\nu < \nu_m < \nu_c$ & ISM  &    \nodata    &  1.9 or 2.9   & \nodata         & Rising spectrum \\
                       & Wind &    \nodata    &               & \nodata         & Rising spectrum \\ 
 $\nu_c < \nu_m < \nu$ & ISM  & $0.8 \pm 0.2$ & $1.9 \pm 0.1$ & \nodata         & $p_{U\!H}$ and $p_X$ are inconsistent \\
                       & Wind &               &               & \nodata         & $p_{U\!H}$ and $p_X$ are inconsistent \\
 $\nu_c < \nu < \nu_m$ & ISM  &    \nodata    & $1.9 \pm 0.1$ & $1/4$           & $\alpha$ does not fit data \\
                       & Wind &    \nodata    &               & $1/4$           & $\alpha$ does not fit data \\ 
 $\nu < \nu_c < \nu_m$ & ISM  &    \nodata    & $1.9 \pm 0.1$ & \nodata         & Rising spectrum \\
                       & Wind &    \nodata    &               & \nodata         & Rising spectrum \\ 
\enddata
\end{deluxetable}
%===== End Table of Predictions =====%     

\begin{figure}
%\figurenum{}
%\epsscale{}
\plotone{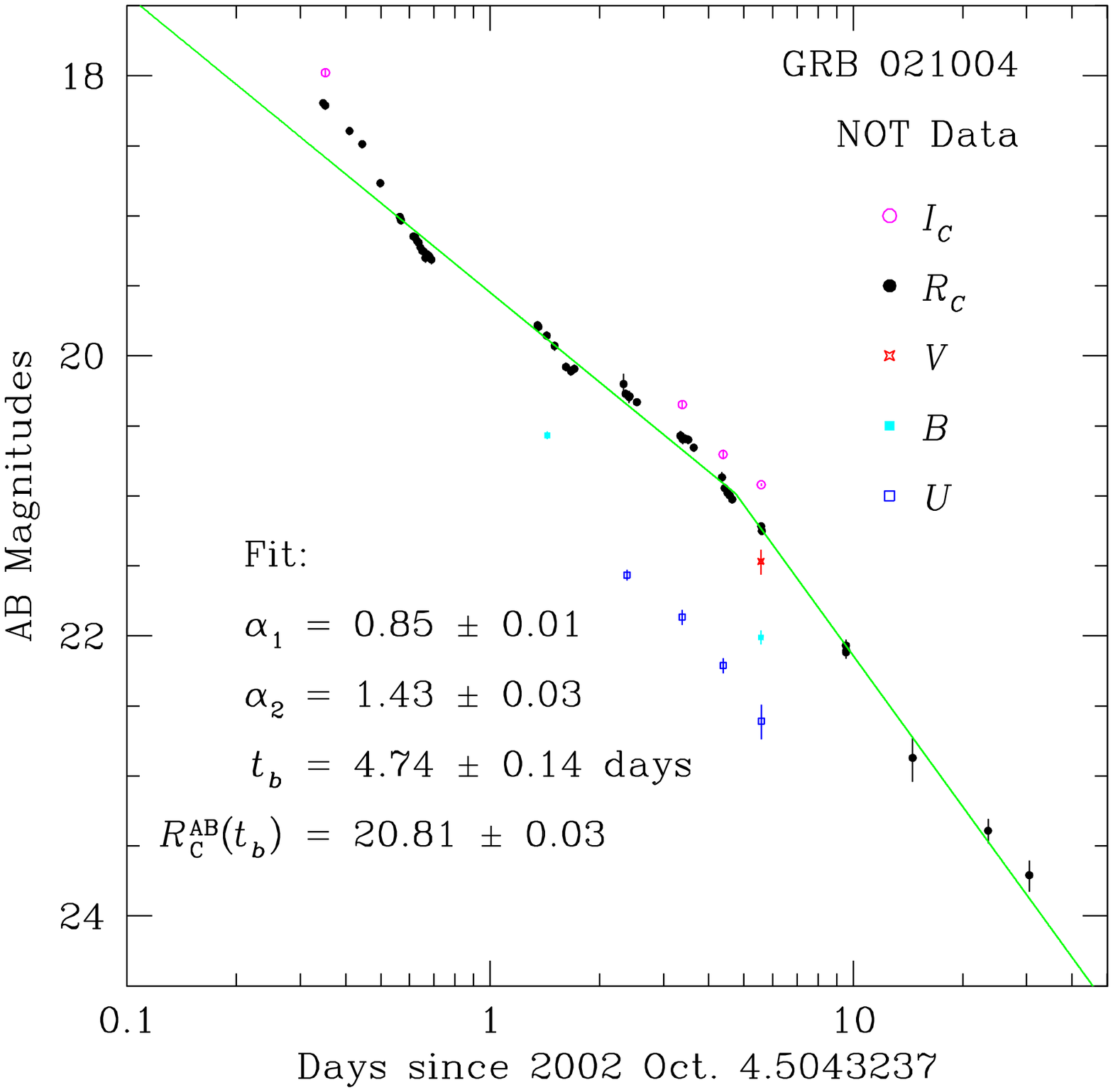} 
\figcaption[./Holland.fig4.ps]{This is our NOT data for
\protect\objectname{GRB~021004}.  The open squares represent the $U$
band, the closed squares represent the $B$ band, the stars represent
the $V$ band, the closed circles represent the $R_C$ band, and the
open circles represent the $I_C$ band.  The line is the best-fitting
broken power law to the $R_C$-band data as described in
\S~\ref{SECTION:light_curve}.  All of the data in this panel have been
scaled to AB magnitudes.\label{FIGURE:not_data}}
\end{figure}

\begin{figure}
%\figurenum{}
%\epsscale{}
\plotone{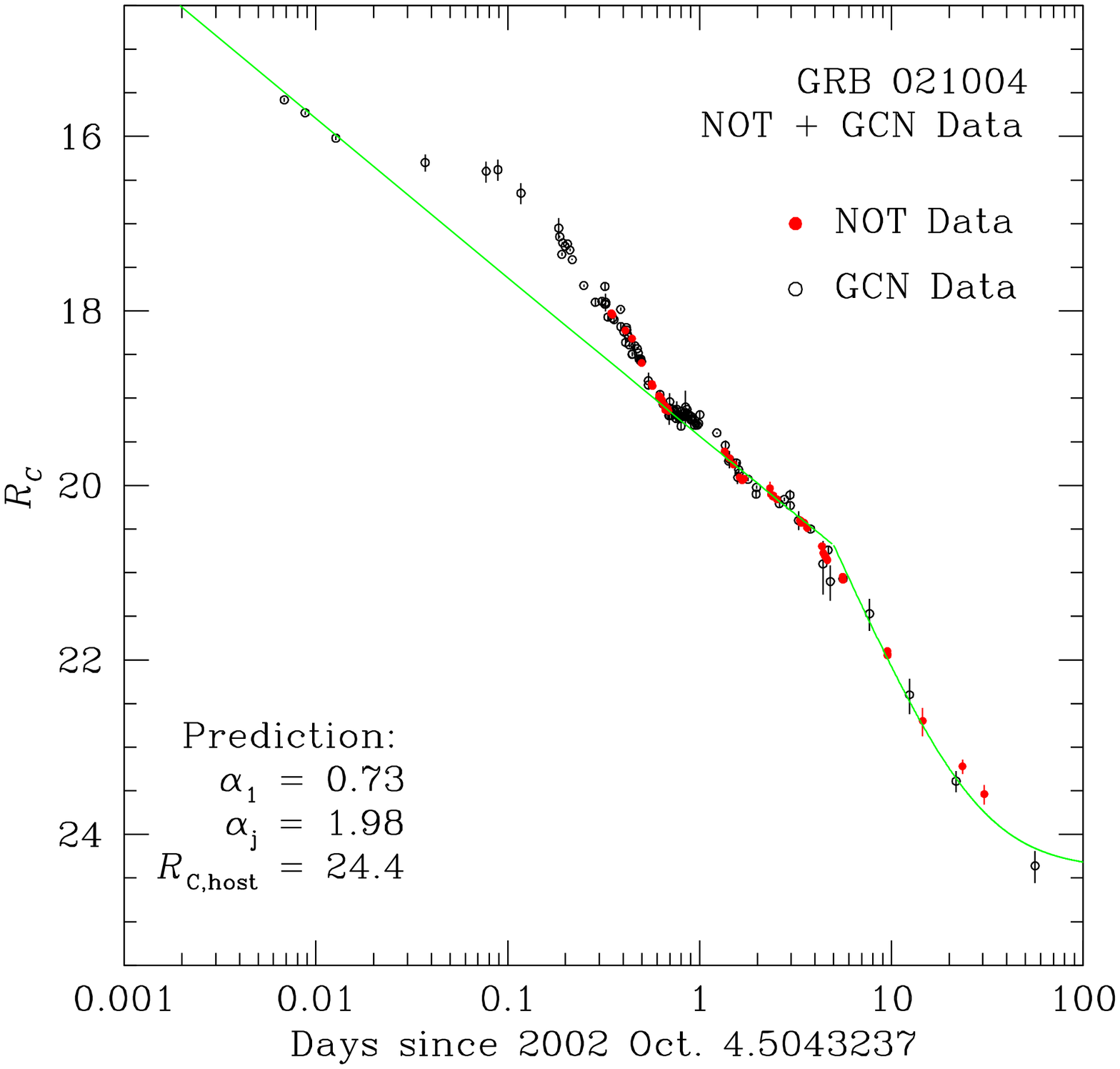}
\figcaption[./Holland.fig5.ps]{This is our NOT $R_C$-band data (closed
circles) and the $R_C$-band data taken from GCN Circulars (open
circles) for \protect\objectname{GRB~021004}.  The later data have
been corrected so that the \citet{F2002} comparison star has $R_C =
15.54 \pm 0.02$~\citep{H2002}.  The line is the predicted (not fitted)
$R_C$-band decay with $p = 2$ in a homogeneous ISM with a host galaxy
that is assumed to have $R_C = 24.4$ \citep{FKM2002}.  The predicted
decay flux has been scaled to the NOT data.\label{FIGURE:gcns}}
\end{figure}

     Fig.~\ref{FIGURE:not_data} shows all of our NOT data for
\objectname{GRB~021004} while Fig.~\ref{FIGURE:gcns} shows our NOT
$R_C$-band data, {\sl and\/} the $R_C$-band data from the GCN
Circulars up to GCN~1717 (2002 Dec.\ 2, \citet{FKM2002}).  The NOT
data constrain the decay slope to be $0.75 \lesssim \alpha \lesssim
1.00$, so we can rule out $\nu_c < \nu < \nu_m$, which predicts
$\alpha = 0.25$.  Therefore, we suggest that $\nu_m < \nu < \nu_c$
during this period and that the electrons were in the slow cooling
regime.  This implies that the cooling frequency is between the
optical and $X$-ray bands between 0.35 and 5.5 days after the burst.
Averaging $p_{U\!H}$ and $p_X$ gives $p = 1.9 \pm 0.1$ (standard
error).  If the burst occurred in a pre-existing stellar wind then the
expected optical decay slope after the cooling break is $\alpha =
1.24$, which is ruled out by the data.  Therefore, we believe that the
burst occurred in an ambient medium that was homogeneous over large
scales and that the cooling frequency is decreasing with time.  In
this scenario the predicted optical decay slope is 0.73 before the
cooling break passes through the optical and 0.98 after.
Fig.~\ref{FIGURE:gcns} shows all of the available $R_C$-band data for
\objectname{GRB~021004} with an optical decay of 0.73 for $t \le 5$
days after the burst.

%===== Begin NOT Light Curve Figure =====%
%===== End NOT Light Curve Figure =====%

%===== Begin GCN Light Curve Figure =====%
%===== End GCN Light Curve Figure =====%

%%%%%%%%%%%%%%%%%%%%%%%%%%%%%%%%%%%%%%%%%%%%%%%%%%%%%%%%%%%%%%%%%%%%%%%%%%%%%%%%%

\subsection{The Optical Light Curve\label{SECTION:light_curve}}

     To search for a break in the optical decay we fit the NOT
$R_C$-band data with a broken power law of the form

\begin{equation}
\label{EQUATION:broken_power_law}
R_{C,\mathrm{fit}}(t) =
   \left \{
      \begin{array}{lll}
         -48.77 - 2.5 \log_{10}\left(f_{\nu}(t_b) {(t/t_b)}^{-\alpha_1}\right), &
         \mathrm{if}  &  t \le t_b \\
         -48.77 - 2.5 \log_{10}\left(f_{\nu}(t_b) {(t/t_b)}^{-\alpha_2}\right), &
         \mathrm{if}  &  t > t_b
      \end{array}
   \right..
\end{equation}

\noindent
where $t_b$ is the time of the break in the power law, $f_{\nu}(t_b)$
is the flux in the $R_C$ band at the time of the break,
$f_\mathrm{host}$ is the $R_C$-band flux from the host galaxy, and
$\alpha_1,\alpha_2$ are the decay slopes before and after the break
respectively.  This formalism makes no assumptions about the physical
causes of the break.  The best fit occurs with $\alpha_1 =
0.85^{+0.01}_{-0.01}$, $\alpha_2 = 1.43^{+0.03}_{-0.03}$, $t_b =
4.74^{+0.14}_{-0.80}$ days, and $f_{\nu}(t_b) =
14.675^{+3.209}_{-0.445}$ $\mu$Jy which corresponds to $R_C(t_b) =
20.81 \pm 0.03$.  This fit is shown in Fig.~\ref{FIGURE:not_data}.
The large formal errors in the break time indicate that the break was
probably gradual and occurred over a period of approximately one day.
 
     The lack of color evolution between 0.35 and 5.5 days means that
the break can not be due to the cooling frequency passing through the
optical.  In addition, the predicted slope after the cooling break is
$\alpha_2 = 0.98$ whereas we find $\alpha_2 = 1.43^{+0.03}_{-0.03}$,
which is inconsistent with the expected slope after the cooling break.
Therefore, the break at $t \approx 5$ days is most likely due to the
Lorentz factor of the fireball falling below $1/\theta_j$ where
$\theta_j$ is the half-opening angle of the jet.  For a jet that is
undergoing sideways expansion the slope of the optical decay after teh
jet break is $\alpha_j = (p + 6)/4$ \citep{DC2001}, so we expect to
see $\alpha_j = 1.98 \pm 0.03$.

%%%%%%%%%%%%%%%%%%%%%%%%%%%%%%%%%%%%%%%%%%%%%%%%%%%%%%%%%%%%%%%%%%%%%%%%%%%%%%%%%

\subsection{The Properties of the Jet\label{SECTION:jet}}

     We applied cosmological $K$ corrections \citep{BFS2001} and
averaged the FREGATE fluences in the 50--300~keV and 7--400~keV bands
to correct them to the 20--2000~keV band.  The corrected fluences were
then averaged to get an isotropic equivalent energy of $E_\mathrm{iso}
= (2.2 \pm 0.3) \times 10^{52}$ erg.  We estimate the opening angle of
the GRB jet using \citet{R1999} and \citet{SPH1999} and the formalism
of \citet{FKS2001}.  This yields $\theta_j \approx (5\fdg8 \pm 0\fdg9)
{(n/0.1)}^{1/8}$ if we assume, as did \citet{FKS2001}, that the
efficiency of converting energy in the ejecta into gamma rays is 0.2.
We note that our result is not very sensitive to this efficiency.
Reducing the efficiency to 0.01 only changes $\theta_j$ by
approximately 30\%.  We estimate that \objectname{GRB~021004}'s
intrinsic energy in gamma rays, after correcting for the jet geometry,
was $E_\gamma \approx (1.1 \pm 0.2) \times 10^{50}
{(n/0.1)}^{1/4}$~erg.  This is only approximately $2 \sigma$ smaller
than the canonical value of $(5 \pm 2) \times 10^{50}$ erg
\citep{FKS2001,PKP2001,PK2002}.

     In order for this burst to have had the ``standard'' energy the
ambient density must be $n \approx 35$ cm$^{-3}$.  This is in
agreement with the particle densities ($0.1 < n < 100$ cm$^{-3}$)
found by \citet{PK2001} for ten GRBs.  It is also similar to the
densities found in some supernova remnants.  \citet{CSM1999} and
\citet{SCM1999} find a mean density of 6~cm$^{-3}$ with large density
gradients for the supernova remnant \objectname{W44} while
\citet{AB1994} find $n \approx 125$ cm$^{-3}$ approximately one pc
from the progenitor of \objectname{SN1978K}.  \citet{LRC2002} and
\citet{NPG2002} find that the observed variations in the optical decay
of \objectname{GRB~021004} are consistent with density variations of
$\Delta n / n \approx 10$ within approximately $10^{17}$--$10^{18}$~cm
of the progenitor.  Therefore, we believe that \objectname{GRB~021004}
occurred in an environment with a mean density that is typical of
other GRBs.

%%%%%%%%%%%%%%%%%%%%%%%%%%%%%%%%%%%%%%%%%%%%%%%%%%%%%%%%%%%%%%%%%%%%%%%%%%%%%%%%%

\section{Discussion\label{SECTION:disc}}

     Our NOT photometry, when combined with the {\sl Chandra\/}
$X$-ray spectrum, suggests that the electron index for
\objectname{GRB~021004} is $p = 1.9 \pm 0.1$.  Most GRBs are well fit
by models with $p \approx 2.3$--2.5 \citep{VKW2000}.  \citet{PK2002}
present models for ten GRBs and find that five are best fit with $p <
2$.  The mean electron index for the ten bursts in their study is
$\overline{p} = 1.9$.  Electron indices of less than two represent
infinite energy in the standard relativistic fireball model
\citep{M2002}.  This can be avoided by introducing an upper limit for
the electron energy distribution \citep{DC2001}, but detailed modeling
of the acceleration of particles in highly relativistic shocks predict
that the electron index should be approximately 2.3 \citep{AGK2001},
which is inconsistent with our results.  The fact that many GRBs
appear to have electron indices of less than two may indicate the need
for detailed hydrodynamic modeling of GRB afterglows in order to
accurately determine the fireball parameters.

     Our results are consistent with a clumpy ISM near the progenitor
as proposed by \citet{LRC2002}.  Their scenario assumes that the GRB
occurred in an ISM that is inhomogeneous on sub-parsec scales and
predicts $p \approx 2$.  Variations in the density of the ambient
medium of $\Delta n / n \approx 10$ can explain the observed
fluctuations on the optical decay of \objectname{GRB~021004}.
\citet{NPG2002} have shown that the deviations from a power law in the
optical decay can be used to reconstruct the density profile in the
vicinity of the progenitor.  Both groups find that the observed
variations in the optical decay of \objectname{GRB~021004} are
consistent with an ambient medium that is homogeneous on scales of
approximately $10^{18}$ cm with a density enhancement of $\Delta n/n
\approx 10$ at approximately $10^{17}$~cm from the progenitor.  The
observations of \objectname{GRB~021004} are also broadly consistent
with a relativistic fireball crossing a discontinuity in the ISM\@.
\citet{DL2002} find that a density jump at $R \approx 10^{17}$ cm from
a GRB can produce an increase of a factor of a few in the optical
flux.

     Extrapolating the predicted pre-cooling break slope to
approximately ten minutes after the burst (see Fig.~\ref{FIGURE:gcns})
shows that the early data of \citet{F2002} are consistent with our
predicted optical decay of $\alpha = 0.73$ before the cooling break.
\citet{KZ2002} present a model where the optical flux before
approximately one hour after the burst is dominated by optical
emission from reverse shocks.  Their model requires $p \approx 2.4$,
which is significantly higher than the electron index which we deduce
from the $X$-ray and optical spectra.  The agreement between the
optical decay which is predicted by the spectral slopes and that seen
at approximately ten minutes after the burst suggests that the
physical mechanism controlling the observed flux at $t \approx 10$
minutes is the same as the one operating at $t > 0.5$ days.

%%%%%%%%%%%%%%%%%%%%%%%%%%%%%%%%%%%%%%%%%%%%%%%%%%%%%%%%%%%%%%%%%%%%%%%%%%%%%%%%%

\section{Conclusions\label{SECTION:conc}}

     We present $U\!BV\!{R_C}{I_C}$ photometry of the OA of
\objectname{GRB~021004} taken at the NOT and {\sl Chandra\/} $X$-ray
data.  The optical data were taken between approximately eight hours
and 30 days after the burst while the $X$-ray data was taken
approximately one day after the burst.  The broad-band optical SED
yields an intrinsic spectral slope of $\beta_{U\!H} = 0.39 \pm 0.12$
while the $X$-ray data gives $\beta_X = 0.94 \pm 0.03$.  There is no
evidence for color evolution between 8.5 hours and 5.5 days after the
burst.  We find an extragalactic extinction of $0.3 \lesssim A_V
\lesssim 0.5$ along the line of sight to the burst.  Our data suggest
that this dust has an SMC extinction law but we are not able to
constrain its redshift.

     The spectral slopes have been combined with the observed
$R_C$-band optical decay to determine that the shocked electrons are
in the slow cooling regime with an electron index of $1.9 \pm 0.1$,
and that the burst occurred in an ISM that is homogeneous on scales
larger than approximately $10^{18}$ cm.  Our data are consistent with
an optical decay of $\alpha = 0.73$ at $t \lesssim 5$ days after the
burst, and $\alpha = 1.98$ after that.  There is evidence that the
transition between the early and late decay slopes occurred over a
period of approximately one day.  This is consistent with a sideways
expanding jet that slows to a Lorentz factor of $\Gamma \approx 10$
approximately six days after the burst.  The total gamma-ray energy in
the burst was $E_\gamma = (1.1 \pm 0.2) \times 10^{50}
{(n/0.1)}^{1/4}$~erg.  The ambient density around
\objectname{GRB~021004} is consistent with what is seen around other
GRBs ($0.1 < n < 100$ cm$^{-3}$) and with densities seen in supernova
remnants \citep{AB1994,CSM1999,SCM1999}.
     
     The rapid localization of \objectname{GRB~021004} and the
near-continuous monitoring of its OA from approximately ten minutes
after the burst occurred has allowed this burst to be studied in
unprecedented detail.  The afterglow shows a large increase in
luminosity approximately 2.5 hours after the burst and a possible
second, smaller increase at $t \approx 1$ day.  Both of these features
would have been missed if optical follow-up had not been immediate and
continuous.  Further, if the OA had not been identified until more
than approximately three hours after the burst the true nature of the
early-time slope would not have been known.  \objectname{GRB~021004}
demonstrates the need for continuous early-time monitoring of GRB OAs.

%%%%%%%%%%%%%%%%%%%%%%%%%%%%%%%%%%%%%%%%%%%%%%%%%%%%%%%%%%%%%%%%%%%%%%%%%%%%%%%%%

\acknowledgements

     We wish to thank the {\sl HETE-II\/} team, Scott Barthelmy, and
the GRB Coordinates Network (GCN) for rapidly providing precise GRB
positions to the astronomical community.  We also wish to thank Arne
Henden for providing precision photometry of stars in the field of
\objectname{GRB~021004}.  STH acknowledges support from the NASA LTSA
grant NAG5--9364.  JPUF gratefully acknowledges support from the
Carlsberg Foundation.  JG acknowledges the receipt of a Marie Curie
Research Grant from the European Commission.  JMCC acknowledges the
receipt of an FPI doctoral fellowship from Spain's Ministerio de
Ciencia y Tecnolog{\'\i}a.  Several of the authors would like to
thank the University of Copenhagen for its hospitality while this
paper was being prepared.  This research has made use of the NASA/IPAC
Extragalactic Database (NED), which is operated by the Jet Propulsion
Laboratory, California Institute of Technology, under contract with
NASA\@.  This paper includes the first data obtained with MOSCA, which
is funded by the Carlsberg Foundation.  This work is supported by the
Danish Natural Science Research Council (SNF).

%%%%%%%%%%%%%%%%%%%%%%%%%%%%%%%%%%%%%%%%%%%%%%%%%%%%%%%%%%%%%%%%%%%%%%%%%%%%%%%%%

%%%%%%%%%%%%%%%%%%%%%%%%%%%%%%%%%%%%%%%%%%%%%%%%%%%%%%%%%%%%%%%%%%%%%%%%%%%%%%%%%
  
\end{document}